\documentclass[12pt]{article}
\usepackage{epsf}
\usepackage{amsmath,amssymb}

\usepackage[dvipdfmx]{graphicx,hyperref}
\usepackage{cite}

\hypersetup{
setpagesize=false,
 bookmarksnumbered=true,%
 bookmarksopen=true,%
 colorlinks=true,%
 linkcolor=blue,
 citecolor=red,
}

\usepackage{comment}
\usepackage{multirow}

\allowdisplaybreaks[1]

\begin{document}

\newcommand{\lsim}{\stackrel{<}{_\sim}}
\newcommand{\gsim}{\stackrel{>}{_\sim}}

\newcommand{\rem}[1]{{$\spadesuit$\bf #1$\spadesuit$}}

\renewcommand{\theequation}{\thesection.\arabic{equation}}

\renewcommand{\thefootnote}{\fnsymbol{footnote}}
\setcounter{footnote}{0}

\begin{titlepage}

\def\thefootnote{\fnsymbol{footnote}}

\begin{center}

\hfill UT-17-05\\
\hfill February, 2017\\

\vskip .75in

{\large \bf 

  Bottom-Tau Unification in Supersymmetric $SU(5)$ Models\\
  with Extra Matters

}

\vskip .75in

{\large 
So Chigusa and Takeo Moroi
}

\vskip 0.25in

\vskip 0.25in

{\em Department of Physics, University of Tokyo,
Tokyo 113-0033, Japan}

\end{center}
\vskip .5in

\begin{abstract}

  We consider $b$-$\tau$ unification in supersymmetric $SU(5)$ grand
 unified theories (GUTs) with extra matters.  The
 renormalization group runnings of $b$ and $\tau$ Yukawa coupling
 constants may be significantly affected by the existence of extra
 matters.  If the extra matters interact with the standard model
 particles (and their superpartners) only through gauge interaction,
 the ratio of the $b$ to $\tau$ Yukawa coupling constants at the GUT
 scale becomes suppressed compared to the case without extra matters.
 This is mainly due to the change of the renormalization group running
 of the $SU(3)_C$ gauge coupling constant.  If the extra matters have
 Yukawa couplings, on the contrary, the (effective) $b$ Yukawa
 coupling at the GUT scale can be enhanced due to the new Yukawa
 interaction.  Such an effect may improve the $b$-$\tau$ unification
 in supersymmetric GUTs.

\end{abstract}

\end{titlepage}

\renewcommand{\thepage}{\arabic{page}}
\setcounter{page}{1}
\renewcommand{\thefootnote}{\#\arabic{footnote}}
\setcounter{footnote}{0}

\section{Introduction}
\label{sec:intro}
\setcounter{equation}{0}

The unification of the standard model (SM) gauge groups into a larger
group, like in $SU(5)$ grand unified theories (GUTs)
\cite{Georgi:1974sy, Georgi:1974yf, Buras:1977yy}, is an attractive
possibility of a new physics beyond the SM.  One of the important
check points of GUTs is the gauge coupling unification which predicts
that the gauge coupling constants of the SM become equal at the
unification scale up to threshold corrections.  It is well known that,
in the SM, there is no strong indication of the unification of the
gauge coupling constants.  In supersymmetric (SUSY) models, on the
other hand, the situation changes because of the existence of the
superparticles as well as up- and down-type Higgses.  In particular,
with the renormalization group equations (RGEs) of the minimal SUSY SM
(MSSM), three gauge coupling constants more-or-less meet at the GUT
scale $M_{\rm GUT}\sim 10^{16}\ {\rm GeV}$ if the mass scale of the
superparticles is $O(1-10)\ {\rm TeV}$ \cite{Dimopoulos:1981yj,
  Einhorn:1981sx, Marciano:1981un, Amaldi:1991cn, Ellis:1990wk,
  Langacker:1991an}.

In simple GUT models based on $SU(5)$, quarks and leptons are embedded
into full multiplets of $SU(5)$.  In particular, the right-handed
down-type quarks and the left-handed lepton doublets are embedded into
the anti-fundamental representations of $SU(5)$, resulting in the
unification of the down-type and charged lepton Yukawa coupling
constants.  In particular, the unification of the Yukawa coupling
constants of $b$-quark and $\tau$-lepton is an interesting check point
of (SUSY) GUTs.  Indeed, the $b$-$\tau$ unification based on SUSY GUTs
has been extensively studied in many literatures \cite{Tobe:2003bc,
  Baer:2012by, Baer:2012cp, Badziak:2012mm, Joshipura:2012sr,
  Elor:2012ig, Baer:2012jp, Anandakrishnan:2012tj, Ajaib:2014ana,
  Miller:2014jza, Gogoladze:2015tfa, Chigusa:2016ody}.

The renormalization group behaviors of the coupling constants are
sensitive to the existence of new particles.  If full multiplets of
$SU(5)$ are added at a single scale, the unification of the gauge
coupling constants is unaffected (at least at the one-loop level),
although the values of the gauge coupling constants depend on the
particle content.  Contrary to the gauge coupling unification, the
unification of the $b$ and $\tau$ Yukawa coupling constants is
expected to be significantly affected by new particles, because the
renormalization group runnings of Yukawa coupling constants are
strongly dependent on the behaviors of the gauge coupling constants.
Importantly, there are various candidates of such new particles, like
new fermions (as well as their superpartners) to realize Peccei-Quinn
symmetry \cite{Peccei:1977hh}, extra chiral superfields in gauge
mediation models \cite{Dine:1993yw, Dine:1994vc, Dine:1995ag}, and so
on.  In addition, existence of extra matters at the mass scale of the
superparticles is required if there exists a non-anomalous discrete
$R$-symmetry \cite{Kurosawa:2001iq, Asano:2011zt}.  Thus, their
effects on the renormalization group runnings of the $b$ and $\tau$
Yukawa coupling constants are of great interest in particular from the
point of view of the Yukawa unification based on SUSY GUTs.

In this paper, we study the $b$-$\tau$ unification in SUSY models with
extra matters which have gauge quantum numbers under the SM gauge
group.  We will see that the existence of the extra matters may
significantly affect the renormalization group running of the $b$ and
$\tau$ Yukawa coupling constants, and hence modify the $b$-$\tau$
unification.  As we will discuss, the ratio of the $b$ to $\tau$
Yukawa coupling constants may become very close to $1$ if the extra
matters have Yukawa couplings with MSSM particles even though in a
large fraction of the parameter space of the MSSM, the Yukawa coupling
constant of $b$ becomes sizably smaller than that of $\tau$ at the GUT
scale.  We will also see that the ratio of the $b$ to $\tau$ Yukawa
coupling constants at the GUT scale becomes smaller in models with
extra matters if they do not have Yukawa interactions.

\section{Model: Brief Overview}
\label{sec:model}
\setcounter{equation}{0}

\begin{table}[t]
  \begin{center} 
    \begin{tabular}{|c|c|c|c|c|}
      \hline
      Classification & Notation ($SU(5)$) & Notation & $SU(5)$ & $G_{\rm SM}$
      \\ \hline
      \multirow{5}{*}{MSSM 3rd generation}
      & \multirow{2}{*}{$\bar{F}$}
      & $b_R^c$ & \multirow{2}{*}{$\bar{\textbf{5}}$}
      & $(\bar{\textbf{3}}, \textbf{1}, \frac{1}{3})$ 
      \\ \cline{3-3} \cline{5-5} 
      & & $l_L$ & & $(\textbf{1}, \textbf{2}, -\frac{1}{2})$ 
      \\ \cline{2-5} 
      & \multirow{3}{*}{$T$}
      & $q_L$ & \multirow{3}{*}{$\textbf{10}$} 
      & $(\textbf{3}, \textbf{2},\frac{1}{6})$ 
      \\ \cline{3-3} \cline{5-5} 
      & & $t_R^c$ & & $(\bar{\textbf{3}}, \textbf{1}, -\frac{2}{3})$
      \\ \cline{3-3} \cline{5-5} 
      & & $\tau_R^c$ & & $(\textbf{1}, \textbf{1}, 1)$
      \\ \hline
      \multirow{10}{*}{Extra Matters} 
      & \multirow{2}{*}{$\bar{F}'$}
      & $D'$
      & \multirow{2}{*}{$\bar{\textbf{5}}$}
      & $(\bar{\textbf{3}}, \textbf{1}, \frac{1}{3})$
      \\ \cline{3-3} \cline{5-5}
      & & $L'$ & & $(\textbf{1}, \textbf{2}, -\frac{1}{2})$
      \\ \cline{2-5}
      & \multirow{3}{*}{$T'$}
      & $Q'$ & \multirow{3}{*}{$\textbf{10}$} 
      & $(\textbf{3}, \textbf{2}, \frac{1}{6})$
      \\ \cline{3-3} \cline{5-5}
      & & $U'$ & & $(\bar{\textbf{3}}, \textbf{1}, -\frac{2}{3})$
      \\ \cline{3-3} \cline{5-5}
      & & $E'$ & & $(\textbf{1}, \textbf{1}, 1)$
      \\ \cline{2-5}
      & \multirow{2}{*}{$F'$}
      & $\bar{D}'$ & \multirow{2}{*}{$\textbf{5}$} 
      & $(\textbf{3}, \textbf{1}, -\frac{1}{3})$
      \\ \cline{3-3} \cline{5-5} 
      & & $\bar{L}'$ & & $(\textbf{1}, \textbf{2}, \frac{1}{2})$
      \\ \cline{2-5}
      & \multirow{3}{*}{$\bar{T}'$} 
      & $\bar{Q}'$ & \multirow{3}{*}{$\overline{\textbf{10}}$} 
      & $(\bar{\textbf{3}}, \textbf{2}, -\frac{1}{6})$
      \\ \cline{3-3} \cline{5-5} 
      & & $\bar{U}'$ & & $(\textbf{3}, \textbf{1}, \frac{2}{3})$
      \\ \cline{3-3} \cline{5-5}
      & & $\bar{E}'$ & & $(\textbf{1}, \textbf{1}, -1)$
      \\ \hline
    \end{tabular}
  \end{center}
  \caption{Notations of chiral superfields used throughout this paper.
    Each column denotes, from left to right, the classification between
    MSSM particles and extra matters, the notation of the $SU(5)$
    multiplet, the notation of the multiplet of the SM gauge group
    $G_{\rm SM} \equiv SU(3)_C \times SU(2)_L \times U(1)_Y$, the
    representation under the unified gauge group $SU(5)$, and the
    representation under $G_{\rm SM}$.}  \label{notations}
\end{table}

We first introduce the model we consider.  The present analysis is
based on \cite{Chigusa:2016ody}, in which $b$-$\tau$ unification in
the MSSM was studied considering proper effective theories.  In the
present study, we include the effects of extra matters into the
analysis of \cite{Chigusa:2016ody}.  We study the model with extra
matters which can be embedded into complete $SU(5)$ representations.
We concentrate on the case where the extra matters are embedded into
$\textbf{5}+\bar{\textbf{5}}$ or $\textbf{10}+\overline{\textbf{10}}$
representations.  The notations for the chiral superfields are
summarized in Tab.\ \ref{notations}.

In the model of our interest, the superpotential can be denoted
as\footnote{Hereafter, the $SU(3)_C$ and $SU(2)_L$ indices are omitted
for notational simplicity.}
\begin{align}
  W = W_{\rm Yukawa} + \mu H_u H_d + W_{\bf 5} + W_{\bf 10},
  \label{Superpot}
\end{align}
with $H_u$ and $H_d$ being the up- and down-type Higgses,
respectively.  Here, $W_{\rm Yukawa}$ is for Yukawa interaction, while
$W_{\bf 5}$ and $W_{\bf 10}$ are SUSY invariant mass terms for extra
matters.  If the Yukawa interactions of the extra matters are
negligible, the relevant part of $W_{\rm Yukawa}$ is given by
\begin{align}
 W_{\rm Yukawa} = y_b H_d q_L b_R^c + y_\tau H_d l_L \tau_R^c + y_t H_u
 q_L t_R^c + \cdots .
 \label{Wyukawa0}
\end{align}
When there exist Yukawa couplings involving extra matters, $W_{\rm
  Yukawa}$ is modified as described below in Sec.\
  \ref{sec:extra_Yukawa}.  We consider $N_{\bf 5}$ pairs of ${\bf
  5}+{\bf \bar{5}}$ (or $N_{\bf 10}$ pairs of ${\bf 10}+\overline{\bf
  10}$), and hence
\begin{align}
  W_{\bf 5} &= \sum_{i=1}^{N_{\bf 5}}
  (\mu_D D_i' \bar{D_i}' + \mu_L L_i' \bar{L_i}'), \\
  W_{\bf 10} &= \sum_{i=1}^{N_{\bf 10}}
  (\mu_Q Q_i' \bar{Q_i}' + \mu_U U_i' \bar{U_i}' +
  \mu_{E} E_i' \bar{E_i}'),
\end{align}
where $i$ is the label of each ${\bf 5}+{\bf \bar{5}}$ or ${\bf
  10}+{\bf \overline{10}}$ pair.  For simplicity, we assume that the
mass parameters for the extra matters in the same standard-model
representations are identical.  Furthermore, the relevant part of the
soft SUSY breaking terms are given by
\begin{align}
  {\cal L}^{\rm (soft)} &= {\cal L}^{\rm (soft)}_{\rm scalar\ mass} +
 {\cal L}^{\rm (soft)}_{\rm trilinear} + \left( - B \mu H_u H_d
 - \frac{1}{2} M_1 \tilde{B} \tilde{B} - \frac{1}{2} M_2 \tilde{W} \tilde{W}
 - \frac{1}{2} M_3 \tilde{g} \tilde{g} + {\rm h.c.} \right) \notag \\
 &\quad + {\cal L}^{\rm (soft)}_{\bf 5} + {\cal L}^{\rm (soft)}_{\bf 10} + \cdots ,
 \label{soft}
\end{align}
where $\tilde{B}$, $\tilde{W}$ and $\tilde{g}$ are Bino, Wino and
gluino, respectively.  (The ``tilde'' is used for SUSY particles.)
Here, ${\cal L}^{\rm (soft)}_{\rm scalar\ mass}$ is soft SUSY breaking
scalar mass terms.  Furthermore, ${\cal L}^{\rm (soft)}_{\rm
trilinear}$ denotes trilinear couplings; when the trilinear couplings
of the extra matters are negligible, it is given by
\begin{align}
 {\cal L}^{\rm (soft)}_{\rm trilinear} =
 - A_b H_d \tilde{q}_L \tilde{b}_R^c 
 - A_\tau H_d \tilde{l}_L \tilde{\tau}_R^c
 - A_t H_u \tilde{q}_L \tilde{t}_R^c + {\rm h.c.} + \cdots .
\end{align}
(The effects of the trilinear couplings of the extra matters are
discussed in Sec.\ \ref{sec:extra_Yukawa}.)  ${\cal L}^{\rm
  (soft)}_{\bf 5}$ and ${\cal L}^{\rm (soft)}_{\bf 10}$ contain
bilinear terms of extra matters,
\begin{align}
 {\cal L}^{\rm (soft)}_{\bf 5} &= \sum_{i=1}^{N_{\bf 5}} (B_D \mu_D \tilde{D}_i'
 \tilde{\bar{D}}_i' + B_L \mu_L \tilde{L}_i' \tilde{\bar{L}}_i') +
 {\rm h.c.}\ , \\
 {\cal L}^{\rm (soft)}_{\bf 10} &= \sum_{i=1}^{N_{\bf 10}} (B_Q \mu_Q \tilde{Q}_i'
 \tilde{\bar{Q}}_i' + B_U \mu_U \tilde{U}_i' \tilde{\bar{U}}_i' +
 B_{E} \mu_{E} \tilde{E}_i' \tilde{\bar{E}}_i') + {\rm h.c.}\ .
\end{align}
As for the SUSY invariant masses of extra matters, we assume that the
SUSY breaking bilinear terms are universal for extra matters with the
same standard-model representations.

Below the mass scale of the SUSY particles, the effective theory
contains only the SM particles (as well as the extra matter fermions
if the SUSY invariant masses of extra matters are smaller than the
masses of SUSY particles).  We denote the Lagrangian of such an
effective theory as
\begin{align}
  {\cal L} =&\ {\cal L}_{\rm kin}^{\rm (SM)}
  + \left( \tilde{y}_b \tilde{H}_{\rm SM} q_L b_R^c +
  \tilde{y}_\tau \tilde{H}_{\rm SM} l_L \tau_R^c + y_t H_{\rm SM} q_L
 t_R^c + {\rm h.c.} \right) \notag \\
 &\quad -  m_{H_{\rm SM}}^2 H_{\rm SM}^\dagger H_{\rm SM}
 - \frac{\lambda}{2} (H_{\rm SM}^\dagger H_{\rm SM})^2
 + {\cal L}^{\rm (extra)} + {\cal L}^{\rm (\tilde{G})} + \cdots ,
  \label{SMlagrangian}
\end{align}
where ${\cal L}_{\rm kin}^{\rm (SM)}$ is the kinetic terms of SM
fields and $H_{\rm SM}$ is the SM-like Higgs doublet (with
$\tilde{H}_{\rm SM}\equiv\epsilon H_{\rm SM}^{*}$).  Yukawa coupling
constants for the effective theories below the mass scale of the SUSY
particles are denoted as $\tilde{y}_b$, $\tilde{y}_\tau$, and
$\tilde{y}_t$.  Furthermore,\footnote{We use same notations for the
  ${\rm SM_{ex}}$ fermions and the corresponding superfields.}
\begin{align}
  {\cal L}^{\rm (extra)} &= {\cal L}_{\rm kin}^{\rm (extra)}
  - \sum_{i=1}^{N_{\bf 5}} (\mu_D \bar{D}_i' D_i' + \mu_L \bar{L}_i'
 L_i' + {\rm h.c.})
  - \sum_{i=1}^{N_{\bf 10}} (\mu_Q \bar{Q}_i' Q_i' + \mu_U \bar{U}_i' U_i'
  + \mu_E \bar{E}_i' E_i' + {\rm h.c.}),
\end{align}
and 
\begin{align}
  {\cal L}^{(\tilde{G})} = {\cal L}_{\rm kin}^{(\tilde{G})} 
  - \frac{1}{2} 
  \left( 
    M_1 \tilde{B} \tilde{B} + M_2 \tilde{W} \tilde{W} + M_3 \tilde{g}
 \tilde{g} + \mbox{h.c.}
  \right),
\end{align}
where ${\cal L}_{\rm kin}^{\rm (extra)}$ and ${\cal L}_{\rm kin}^{\rm
  (\tilde{G})}$ are kinetic terms of extra matter fermions and
gauginos, respectively.  In Eq.\ \eqref{SMlagrangian}, ${\cal L}^{\rm
  (extra)}$ and ${\cal L}^{\rm (\tilde{G})}$ should be omitted below
the mass scale of the extra matter fermions and gauginos,
respectively.

Some of the Lagrangian parameters are related to each other at the GUT
scale $M_{\rm GUT}$.  (In our analysis, we define $M_{\rm GUT}$ as the
scale at which $U(1)_Y$ and $SU(2)_L$ gauge coupling constants become
equal.)  For simplicity, we assume that the SUSY breaking scalar mass
parameters are degenerate at the GUT scale for scalars with same
$SU(5)$ representations.  For the bilinear terms and the soft SUSY
breaking parameters, we neglect the threshold corrections at the GUT
scale.  Then, we parametrize the Lagrangian parameters at $Q=M_{\rm
GUT}$ (with $Q$ being the renormalization scale) as
\begin{align}
 & \mu_{D} (M_{\rm GUT}) = \mu_{L} (M_{\rm GUT}) \equiv \mu_{\bf 5}, \\
 & \mu_{Q} (M_{\rm GUT}) = \mu_{U} (M_{\rm GUT}) = \mu_{E} (M_{\rm
 GUT}) \equiv \mu_{\bf 10} \\ &
 m_{\tilde{D}}^2 (M_{\rm GUT}) = m_{\tilde{L}}^2 (M_{\rm GUT}) \equiv
 m_{\bf \bar{5}}^2, \\ &
 m_{\tilde{D'}}^2 (M_{\rm GUT}) = m_{\tilde{L'}}^2 (M_{\rm GUT}) =
 m_{\bf \bar{5}}^2, \\ &
 m_{\tilde{Q}}^2 (M_{\rm GUT}) = m_{\tilde{U}}^2 (M_{\rm GUT}) =
 m_{\tilde{E}}^2 (M_{\rm GUT}) \equiv m_{\bf 10}^2, \\ &
 m_{\tilde{Q'}}^2 (M_{\rm GUT}) = m_{\tilde{U'}}^2 (M_{\rm GUT}) =
 m_{\tilde{E'}}^2 (M_{\rm GUT}) = m_{\bf 10}^2, \\ &
 m_{H_u}^2 (M_{\rm GUT}) \equiv m_{H{\bf 5}}^2, \\ &
 m_{H_d}^2 (M_{\rm GUT}) \equiv m_{H\bar{\bf 5}}^2, \\ &
 A_b (M_{\rm GUT}) = A_\tau (M_{\rm GUT}) \equiv a_d,\\ &
 A_t (M_{\rm GUT}) \equiv a_u, \\ &
 B_D (M_{\rm GUT}) = B_L (M_{\rm GUT}) = m_{\bf \bar{5}}, \\ &
 B_Q (M_{\rm GUT}) = B_U (M_{\rm GUT}) = B_E (M_{\rm GUT}) = m_{\bf 10},  
\end{align}
where $m_{\tilde{D}}^2$, $m_{\tilde{L}}^2$, $m_{\tilde{Q}}^2$,
$m_{\tilde{U}}^2$ and $m_{\tilde{E}}^2$ are soft SUSY breaking mass
squared parameters of $\tilde{b}_R^c$, $\tilde{l}_L$, $\tilde{q}_L$,
$\tilde{t}_R^c$, and $\tilde{\tau}_R^c$, respectively.  In addition,
we impose the same boundary condition for $m_{\tilde{\bar{\Phi}}'}^2$
as $m_{\tilde{\Phi}'}^2$ ($\Phi = D,L,Q,U,E$).  For gaugino masses, we
adopt the simple GUT relation:
\begin{align}
 M_1 (M_{\rm GUT}) = M_2 (M_{\rm GUT}) = M_3 (M_{\rm GUT}) \equiv m_{1/2}.
\end{align}

The mass spectrum of the SUSY particles (including those in the extra
matter sector) is determined by solving the RGEs with the boundary
conditions given above.  Importantly, the masses of the scalars are
comparable to or larger than the gaugino masses in the model of our
interest because of the renormalization group effects.  We also
comment here that the gaugino masses may be much smaller than the
scalar masses, as suggested by several models of SUSY breaking
\cite{Ibe:2006de, Ibe:2011aa, ArkaniHamed:2012gw}.  Thus, we do not
exclude the possibility that the gaugino masses are hierarchically
smaller than the scalar masses.

\begin{table}[t]
  \begin{center}
    \begin{tabular}{|c||c|c|}
      \hline
      Effective theories & Particle content & RGEs \\ \hline
      SM & SM particles & two-loop \\ \hline
      ${\rm SM_{ex}}$ & SM particles and extra fermions & two-loop \\ \hline
      ${\rm \tilde{G} SM}$ & SM particles and gauginos & two-loop \\ \hline
      ${\rm \tilde{G} SM_{ex}}$ & SM particles, gauginos and extra fermions & two-loop \\ \hline
      MSSM & MSSM particles & two-loop \\ \hline
      ${\rm MSSM_{ex}}$ & MSSM particles and extra matters 
      & two-loop \\ \hline
    \end{tabular}
    \caption{Effective theories used in our analysis.}
    \label{tab:effectiveth}
  \end{center}
\end{table}
  
\begin{table}[t]
 \begin{center}
  \begin{tabular}{|c||c|c|c|}
   \hline
   Effective theories & $M_{\rm ex} < M_{\tilde{G}} < M_S$ &
	   $M_{\tilde{G}} < M_{\rm ex} < M_S$ & $M_{\tilde{G}} < M_S <
	       M_{\rm ex}$ \\ \hline
   SM & $m_t < Q < M_{\rm ex}$ & $m_t < Q < M_{\tilde{G}}$ & $m_t < Q <
	       M_{\tilde{G}}$ \\ \hline
   ${\rm SM_{ex}}$ & $M_{\rm
       ex} < Q < M_{\tilde{G}}$ & & \\ \hline
   ${\rm \tilde{G} SM}$ & & $M_{\tilde{G}} < Q < M_{\rm ex}$ &
	       $M_{\tilde{G}} < Q < M_S$ \\ \hline
   ${\rm \tilde{G} SM_{ex}}$ & $M_{\tilde{G}} < Q < M_S$ & $M_{\rm ex} <
	   Q < M_S$ & \\ \hline
   MSSM & & & $M_S < Q < M_{\rm ex}$ \\ \hline
   ${\rm MSSM_{ex}}$ & $M_S < Q < M_{\rm GUT}$ & $M_S < Q < M_{\rm GUT}$
	   & $M_{\rm ex} < Q < M_{\rm GUT}$ \\ \hline
  \end{tabular}
  \caption{Effective theories for each renormalization scale $Q$.}
  \label{tab:effective}
 \end{center}
\end{table}

In order to calculate the renormalization group running of coupling
constants from the weak scale to the GUT scale $M_{\rm GUT}$, we
consider several effective theories; particle contents of all the
effective theories used in our analysis are summarized in Tab.\
\ref{tab:effectiveth}.  In the present analysis, there are three
important mass scales, i.e., the gaugino mass scale $M_{\tilde{G}}$,
the sfermion mass scale $M_S$, and the extra fermion mass scale
$M_{\rm ex}$, at which the effective theory changes from one to
another.  As we have mentioned, the scalar masses are comparable to or
larger than the gaugino masses, and hence $M_{\tilde{G}}<M_S$.  In
addition, the masses of the scalars in the extra matter sector have
two contributions, i.e., SUSY invariant mass parameters and soft SUSY
breaking masses (denoted as $\mu_\Phi$ and $m_{\tilde{\Phi}}^2$,
respectively); the scalar masses in the extra matter sector is
$\sim\sqrt{\mu_\Phi^2+m_{\tilde{\Phi}}^2}$.  Thus, we assume that the
scalars in the extra matter sector decouple from the effective theory
at the renormalization scale $Q=\mbox{max} (M_S,M_{\rm ex})$.  The
relevant effective theory for each renormalization scale depends on
$M_{\tilde{G}}$, $M_S$, and $M_{\rm ex}$, as summarized in Tab.\
\ref{tab:effective}.

For all the effective theories mentioned above, we use two-loop RGEs;
for SUSY models, we use the Susyno package \cite{Fonseca:2011sy},
while the RGEs for non-SUSY theories are calculated based on
\cite{Machacek:1983tz, Machacek:1983fi, Machacek:1984zw}.  In
addition, at each energy threshold, one effective theory is matched to
another, taking into account one-loop threshold corrections to
Lagrangian parameters.  In the following, we summarize the important
effects.

At the sfermion mass scale $M_S$, two Higgs doublets in ${\rm
MSSM_{(ex)}}$ are matched to the SM-like Higgs as
\begin{align}
 H_{\rm SM} = H_u \sin \beta + H_d^{*} \cos \beta,
\end{align}
where $\tan \beta$ is the ratio of the vacuum expectation value of
$H_u^0$ to that of $H_d^0$.  The boundary condition for the Higgs
quartic coupling $\lambda$ at $M_S$ is
\begin{align}
 \lambda (M_S) = \frac{g_1^2 (M_S) + g_2^2 (M_S)}{4} \cos^2 2\beta +
 \delta \lambda,
\end{align}
where $\delta \lambda$ is the threshold correction due to heavy scalar
particles (in particular, stops).  In addition, the mass of the
pseudo-scalar Higgs, which is a component of the heavy Higgs multiplet
$H_{\rm heavy} = H_u \cos \beta - H_d^{*} \sin \beta$, is determined at
this scale as
\begin{align}
 m_A^2 = [m_{H_u}^2 + m_{H_d}^2 + 2\mu^2 - m_{H_{\rm SM}}^2]_{Q = M_S},
\end{align}
where $\mu^2$ is determined from the following radiative electroweak
symmetry breaking condition:
\begin{align}
 \mu^2 &= - m_{H_{\rm SM}^2} - m_{H_u}^2 \sin^2 \beta - m_{H_d}^2 \cos^2
 \beta + B \mu \sin 2\beta,\label{muSq}\\
 B \mu &= \frac{1}{2} (m_{H_u}^2 - m_{H_d}^2) \tan 2\beta.
\end{align}

The Yukawa coupling constants $y_f$ (with $f=t$, $b$, and $\tau$) are
matched to $\tilde{y}_f$ at $Q=M_S$, using the mixing angle $\beta$.
In our analysis, the threshold correction to the bottom Yukawa
coupling constant at $Q = M_S$ is important.  The correction
$\Delta_b$ is defined by
\begin{align}
  \tilde{y}_b (M_S) = y_b (M_S) \cos \beta (1 + \Delta_b),
\end{align}
where $\tilde{y}_b$ and $y_b$ are the bottom quark Yukawa coupling
constants in the effective theory used just below and just above
$M_S$, respectively.  The most important contributions to $\Delta_b$
come from the sbottom-gluino and stop-chargino diagrams
\cite{Hall:1993gn, Carena:1994bv, Blazek:1995nv}; at the leading order
of the mass-insertion approximation, these contributions are given by
\begin{align}
  \Delta_b \simeq
  \left[
    \frac{g_3^2}{6\pi^2} M_3 I(m_{\tilde b_1}^2,m_{\tilde b_2}^2,M_3^2)
    + \frac{y_t}{16\pi^2} A_t I(m_{\tilde t_1}^2,m_{\tilde t_2}^2,\mu^2)
 \right] \mu \tan\beta,
 \label{eq:deltab}
\end{align}
where $m_{\tilde b_1}$ and $m_{\tilde t_1}$ ($m_{\tilde b_2}$ and
$m_{\tilde t_2}$) are masses of lighter (heavier) stop and sbottom,
respectively, and
\begin{align}
  I(a,b,c) = -\frac{ab\ln (a/b)+bc\ln (b/c)+ca\ln (c/a)}{(a-b)(b-c)(c-a)}.
\end{align}
(In our numerical analysis, we use the full one-loop expression of
$\Delta_b$.)  The important point is that $\Delta_b$ is approximately
proportional to $\mu \tan \beta$, resulting in the large correction to
the bottom Yukawa coupling constant in the models with heavy Higgsinos
or those with large $\tan \beta$.

We also include threshold corrections to the Wino and Bino masses at $Q
= M_S$ due to the Higgs-Higgsino loop diagram \cite{Giudice:1998xp}:
\begin{align}
  \delta M_1 = \frac{g_1^{2} (M_{\rm S})}{16 \pi^{2}} L, ~~~
  \delta M_2 =&\,  \frac{g_2^{2} (M_{\rm S})}{16 \pi^{2}} L,
\end{align}
where
\begin{eqnarray}
 L \equiv \mu \sin2\beta 
  \frac{m_{A}^{2}}{\mu^{2}-m_{A}^{2}} \ln \frac{\mu^{2}}{m_{A}^{2}}.
\end{eqnarray}

At $Q = M_{\tilde{G}}$ and $Q = M_{\rm ex}$, we take into account
one-loop threshold corrections to gauge coupling constants, gaugino
masses, and scalar masses due to loop diagrams involving gauginos and
extra matters.  Then, at $Q = m_t$ the SM-like Higgs mass is evaluated
as
\begin{align}
 m_h^2 = 2 \lambda(m_t) v^2 + \delta m_h^2,
\end{align}
where $v \simeq 174\ {\rm GeV}$ is the vacuum expectation value of the
SM-like Higgs boson and $\delta m_h^2$ is the threshold correction.

\section{Numerical Results}
\label{sec:numerical}
\setcounter{equation}{0}

In this section, we show the results of our numerical study.  In
addition to the SM parameters, the present model contains ten new
parameters, $\tan\beta$, $m_{\bf \bar{5}}^2$, $m_{\bf 10}^2$,
$m_{H{\bf{5}}}^2$, $m_{H{\bf \bar{5}}}^2$, $m_{1/2}$, $\mu$, $B$,
$\mu_{\bf 5}$, and $\mu_{\bf 10}$, ignoring the Yukawa and the
trilinear couplings related to the extra matters.  Among them, $\mu$
and $B$ are determined at the sfermion mass scale $M_S$ to fix the
vacuum expectation value of the SM-like Higgs boson $v$ and
$\tan\beta$.

We numerically solve RGEs from the weak scale to the GUT scale.  Our
numerical calculation is based on the SOFTSUSY package
\cite{Allanach:2001kg}, in which three-loop RGEs for the effective
theory below the electroweak scale and two-loop RGEs for the MSSM are
implemented.  We have implemented into SOFTSUSY package two-loop RGEs
for the other effective theories listed in Tab.\ \ref{tab:effective},
i.e., SM, ${\rm SM_{ex}}$, $\tilde{G}$SM, ${\rm \tilde{G}SM_{ex}}$,
and ${\rm MSSM_{ex}}$.  In addition, one-loop threshold corrections
due to the diagrams with SUSY particles or extra matters in the loop
are included at relevant thresholds.  In our numerical calculation,
$M_{\rm S}$ is taken to be the geometric mean of the stop masses,
while we take $M_{\tilde{G}}=M_3$.  $M_{\rm ex}$ is set to the mass of
the bottom-like extra fermion mass $\mu_D$ for models with $N_{\bf 5}
> 0$, and is set to the geometric mean of top-like extra fermion
masses for models with $N_{\bf 10} > 0$.  The gauge and Yukawa
coupling constants are determined based on \cite{Agashe:2014kda}.  In
particular, we use the bottom quark mass of $m_b^{(\overline{\rm
    MS})}(m_b)=4.18\ {\rm GeV}$, the top quark mass of $m_t=173.21\
{\rm GeV}$, and $\alpha_{3} (M_Z)=0.1185$ (with
$\alpha_{3}=g_3^2/4\pi$).

\subsection{Extra matters without Yukawa couplings}
\label{sec:extra_noYukawa}

Let us now study the effects of extra matters on the $b$-$\tau$
unification in SUSY GUT.  We first consider the case where the extra
matters interact with the MSSM particles only through gauge
interactions.

Because the boundary conditions for the Yukawa coupling constants are
fixed by using the fermion masses, $y_b(M_{\rm GUT})$ and $y_{\tau}
(M_{\rm GUT})$ may differ in the present analysis.  To quantify the
difference, we define
\begin{align}
  R_{b\tau} \equiv \frac{y_b (M_{\rm GUT})}{y_\tau (M_{\rm GUT})}.
\end{align}
We calculate $R_{b\tau}$ as a function of model parameters, and study
how it is affected by extra matters.  If there is no source of the GUT
scale threshold corrections other than the splitting of the masses of
GUT scale particles, then $(R_{b\tau} - 1) \sim O(1) \%$.  Thus, if
$R_{b\tau}$ is (much) larger than $\sim O(1) \%$, it indicates a
sizable threshold correction at the GUT scale and/or a non-trivial
flavor physics at the GUT scale or below.

If the Yukawa interactions of the extra matters are negligible, the
main effect of extra matters on the Yukawa unification is through the
enhancement of the gauge coupling constants at high energy scale.
With extra matters, the coupling constants of $SU(3)_C \times SU(2)_L
\times U(1)_Y$ become larger.  This can be understood by examining the
RGEs of the gauge coupling constants; in SUSY models, they are given
by
\begin{align}
  \frac{d}{d \ln Q} g_a &= \frac{g_a^3}{16\pi^2}
  \left[b_a + (N_{\bf 5} + 3 N_{\bf 10})\right] + \cdots,
\end{align}
where $(b_1,b_2,b_3)=(\frac{33}{5}, 1, -3)$, and $\cdots$ in the above
equation denotes higher order effects.  One can see that, with
non-vanishing $N_{\bf 5}$ or $N_{\bf 10}$, the beta-function
coefficients become larger, resulting in the enhancement of the gauge
coupling constants at higher scale.  In particular, the enhancement of
$g_3$ is the most important because of the largeness of the coupling
constant $g_3$ itself.  The enhanced gauge coupling constants affect
the renormalization group running of $y_b$ and $y_\tau$, whose RGEs
are given by
\begin{align}
  \frac{d}{d \ln Q} y_b &= \frac{y_b}{16\pi^2}
  \left( 
    y_t^2 + 6 y_b^2 + y_\tau^2 - \frac{7}{15} g_1^2
    - 3 g_2^2 - \frac{16}{3} g_3^2 
  \right) + \cdots,
  \label{RGEyb}
  \\
  \frac{d}{d \ln Q} y_\tau &= \frac{y_\tau}{16\pi^2} 
  \left(
    3 y_b^2 + 4  y_\tau^2 - 
    \frac{9}{5} g_1^2 - 3 g_2^2
  \right) + \cdots,
  \label{RGEytau}
\end{align}
where the mixings between different generations are neglected.  With
the low-scale values of the Yukawa coupling constants being fixed to
realize the observed fermion masses, the above equations indicate that
the Yukawa coupling constants at $M_{\rm GUT}$ is more suppressed as
the gauge coupling constants become larger.  Due to this effect, $y_b$
is more suppressed than $y_\tau$ because $g_3$ only affects the
running of $y_b$.

\begin{figure}[t]
 \begin{minipage}{0.48\hsize}
  \begin{center}
   \includegraphics[width=0.98\hsize]{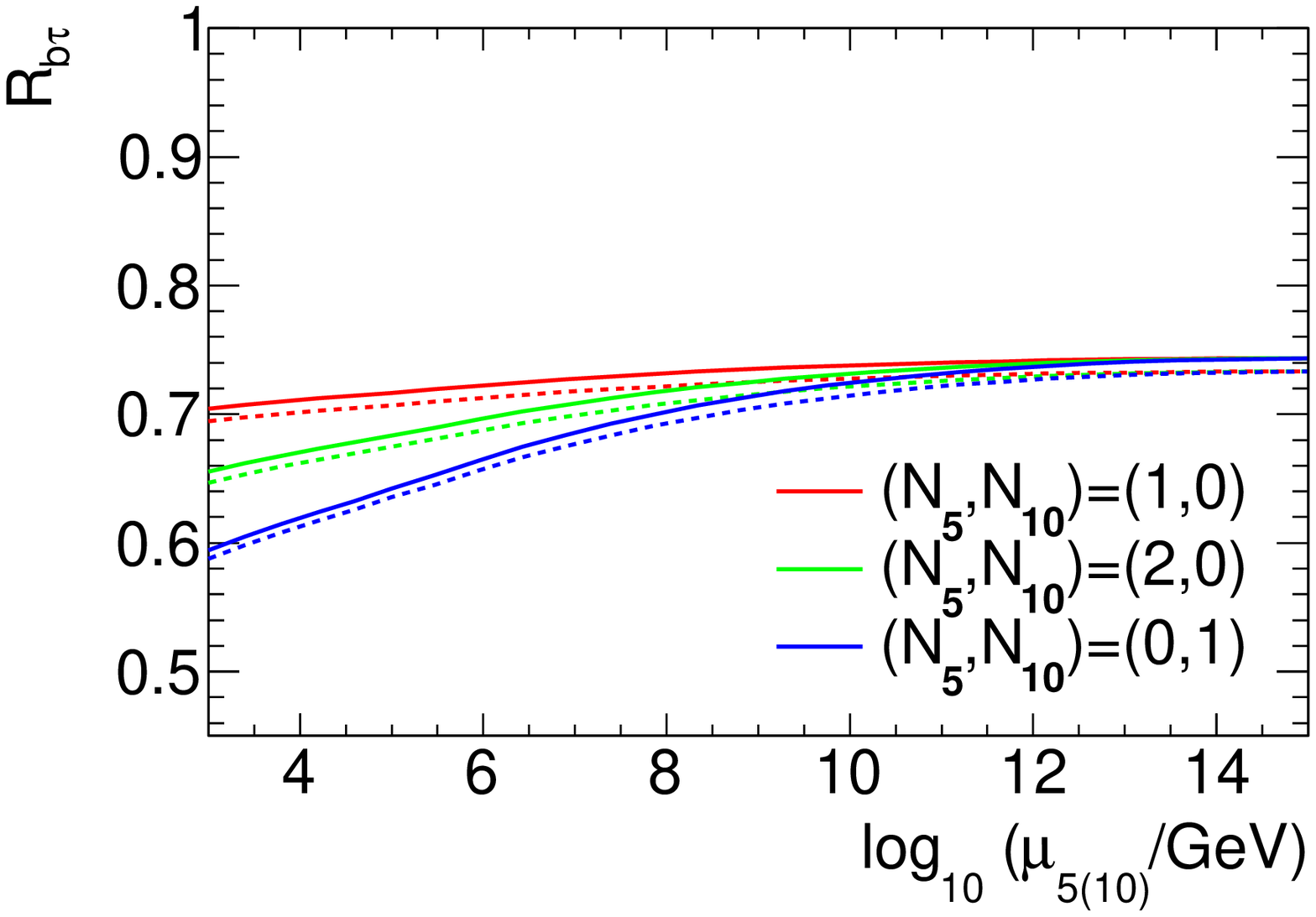}
  \end{center}
 \end{minipage}
 \begin{minipage}{0.48\hsize}
  \begin{center}
   \includegraphics[width=0.98\hsize]{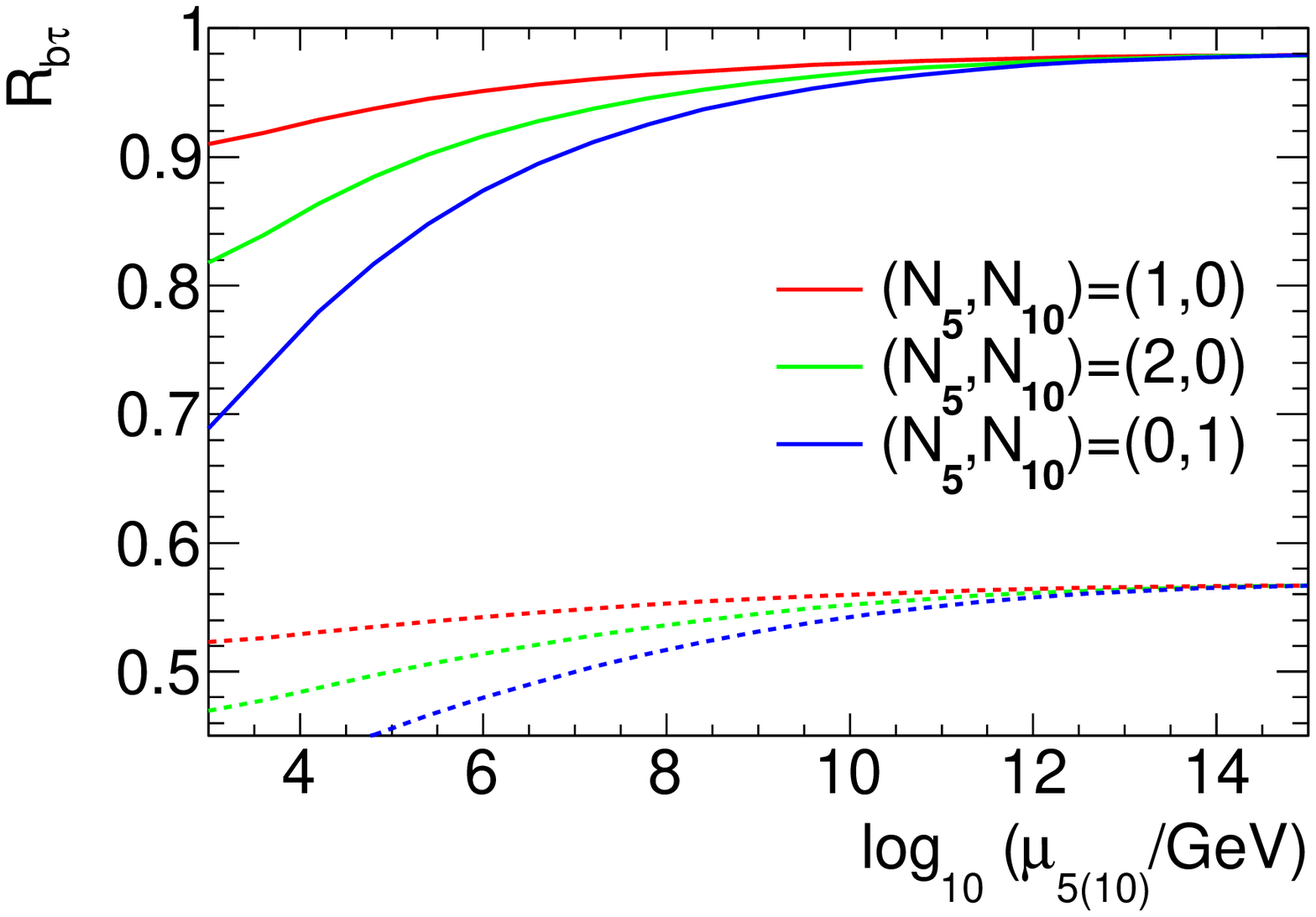}
  \end{center}
 \end{minipage}
 \caption{$R_{b\tau}$ as a function of the mass scale of the extra
   matters for models with low $\tan \beta$ (left) and high $\tan
   \beta$ (right), taking $(N_{\bf 5},N_{\bf 10}) = (1,0)$ (red),
   $(2,0)$ (green), and $(0,1)$ (blue).  The horizontal axis denotes
   the value of $\mu_{\bf 5}$ for red and green lines and that of
   $\mu_{\bf 10}$ for blue lines.  We consider both signs of the SUSY
   invariant Higgs mass: $\mu > 0$ (dotted) and $\mu < 0$ (solid).
   The boundary conditions used in the left figure are $m_{\bf
     \bar{5}} = m_{\bf 10} = m_{1/2} = 100\ {\rm TeV}$, $m_{H{\bf 5}}
   = m_{H{\bf \bar{5}}} = 80\ {\rm TeV}$, and $a_d = a_u = 0$.  $\tan
   \beta$ is determined so that $m_h = 125.09\ {\rm GeV}$, which
   results in $2.9 < \tan \beta < 3.1$.  The boundary conditions for
   the right figure are $\tan \beta = 50$, $m_{\bf \bar{5}} = m_{\bf
     10} = m_{H{\bf 5}} = m_{H{\bf \bar{5}}} = 0$, and $a_d = a_u =
   0$.  $m_{1/2}$ is determined so that $m_h = 125.09\ {\rm GeV}$,
   which results in $4.5\ {\rm TeV} < m_{1/2} < 6.5\ {\rm TeV}$.}
 \label{MexVsR}
\end{figure}

In Fig.\ \ref{MexVsR}, in order to investigate how these effects
affect $R_{b\tau}$, we show $R_{b\tau}$ as a function of the mass
scale of extra matters.  The red, green, and blue lines correspond to
the models with $(N_{\bf 5},N_{\bf 10}) = (1,0)$, $(2,0)$, and
$(0,1)$, respectively.  (Thus, the horizontal axis corresponds to
$\mu_{\bf 5}$ for red and green lines and $\mu_{\bf 10}$ for blue
lines.)  The dotted and solid lines are for models with $\mu > 0$ and
$\mu < 0$, respectively.  For the left figure, we take mSUGRA-like
boundary conditions, $m_{\bf \bar{5}} = m_{\bf 10} = m_{1/2} = 100\
{\rm TeV}$, $m_{H{\bf 5}} = m_{H{\bf \bar{5}}} = 80\ {\rm TeV}$, and
$a_d = a_u = 0$.  Here, $\tan \beta$ is determined so that the SM-like
Higgs mass is given by the observed value $m_h = 125.09\ {\rm GeV}$;
then, it takes values in the range $2.9 < \tan \beta < 3.1$.  The
right figure shows the results for the model with gaugino mediation
boundary conditions \cite{Kaplan:1999ac, Chacko:1999mi}, taking $\tan
\beta = 50$, $m_{\bf \bar{5}} = m_{\bf 10} = m_{H{\bf 5}} = m_{H{\bf
\bar{5}}} = 0$, and $a_d = a_u = 0$. In this case, $m_{1/2}$ is tuned
so that $m_h$ is equal to the observed Higgs mass, which gives $4.5\
{\rm TeV} < m_{1/2} < 6.5\ {\rm TeV}$.

As can be easily understood, the effects of extra matters on the
runnings of $y_b$ and $y_\tau$ are more enhanced as the masses of the
extra matters become smaller.  We can see that $R_{b\tau}$ is
suppressed by $\sim 10\ \%$ when the mass scale of the extra matters
is at around the TeV scale, while $R_{b\tau}$ approaches to the MSSM
value when the mass scale $M_{\rm ex}$ becomes close to the GUT scale.
In the MSSM, it is often the case that $R_{b\tau}$ is significantly
smaller than 1 in particular when $\tan \beta$ is small.  As we have
seen, the effects of extra matters make $R_{b\tau}$ smaller if extra
matters interact with MSSM particles only through gauge interactions.
We also note here that the deviation between a solid line and the
corresponding dotted line approximately shows (twice) the size of the
threshold correction $\Delta_b$, since the sign of $\Delta_b$ is
determined by that of $\mu$.  The size of $\Delta_b$ is also affected
by the change of $M_{\rm ex}$ because the mass spectrum of the MSSM
particles also depends on $M_{\rm ex}$.  With the model parameters for
the solid lines in Fig.\ \ref{MexVsR} (right), for example,
$|\Delta_b|$ is enhanced as $M_{\rm ex}$ becomes smaller.  However,
the suppression of $R_{b\tau}$ due to the enhancement of $g_3$ at
higher scale is more significant; consequently, $R_{b\tau}$ becomes
smaller as $M_{\rm ex}$ decreases as shown in Fig.\ \ref{MexVsR}.

\subsection{Extra matters with Yukawa couplings}
\label{sec:extra_Yukawa}

So far, we have considered the case where the Yukawa interactions of
the extra matters are negligibly small.  However, the extra matters
may couple to MSSM chiral multiplets via Yukawa couplings.  With such
new interaction, the renormalization group runnings of the coupling
and mass parameters may be changed, affecting the unification of
Yukawa coupling constants.  In this subsection, we show that this is
indeed the case.  Among several possibilities, we introduce the Yukawa
interaction for the extra matter in ${\bf 10}$ representation of
$SU(5)$.  We will see that, in such a model, the extra matters may
help to make the $b$-$\tau$ unification successful.

Here, we consider the case with $(N_{\bf 5},N_{\bf 10}) = (0,1)$, and
study the effect of the Yukawa interactions of extra matters.  We
concentrate on the case where the extra matter scale is comparable to
or higher than $M_{\rm S}$ so that all the extra matters (i.e.,
fermions and scalars) simultaneously decouple from the effective
theory at a single scale $M_{\rm ex}$.

For the study of such a case, it is instructive to use the fact that
the Yukawa interaction above the GUT scale can be written in the
following form:
\begin{align}
  W_{\rm Yukawa}^{(SU(5))} = \eta_{b,\tau} \bar{\bf 5}_H {\bf 10}_Y \bar{F} 
  + 
  {\bf 5}_H 
  \left( \begin{array}{cc}
      {\bf 10}_Y & {\bf 10}_0
    \end{array} \right)
  \left( \begin{array}{cc}
      \eta_t & \eta_t^{(1)} \\
      \eta_t^{(1)} & \eta_t^{(2)} \\
    \end{array} \right)
  \left( \begin{array}{c}
      {\bf 10}_Y \\
      {\bf 10}_0 \\
    \end{array} \right),
\end{align}
where $\eta$'s are coupling constants.  Here, ${\bf 10}_Y$ and ${\bf
  10}_0$ are chiral multiplets in the ${\bf 10}$ representation of
$SU(5)$, and are given by linear combinations of $T$ and $T'$.  Notice
that, in this basis, only ${\bf 10}_Y$ couples to $\bar{F}$ though the
Yukawa interaction.  In addition, ${\bf 5}_H$ and $\bar{\bf 5}_H$ are
chiral multiplets containing up- and down-type Higgses, respectively.
In order to make our point clearer, we take
$\eta_t^{(1)}=\eta_t^{(2)}=0$ in the following analysis.  With such an
assumption, the Yukawa interaction below the GUT scale is given by
\begin{align}
  W_{\rm Yukawa} = &\,
  y'_b H_d Q_Y b_R^c 
  + y'_\tau H_d l_L E_Y
  + y'_t H_u Q_Y U_Y,
  \label{W_yukawa}
\end{align}
where $Q_Y$, $U_Y$, and $E_Y$ are chiral superfields embedded into
${\bf 10}_Y$.  (Chiral multiplets embedded into ${\bf 10}_0$ are
denoted as $Q_0$, $U_0$, and $E_0$.)  Denoting the SUSY invariant mass
terms for the extra matters as
\begin{align}
  W_{\bf 10} = \sum_{\Phi=Q,U,E}
  (\mu_{\Phi_Y} \Phi_Y \bar{\Phi}' + \mu_{\Phi_0} \Phi_0 \bar{\Phi}'),
\end{align}
we obtain
\begin{align}
  \left( \begin{array}{c}
      Q_Y \\
      Q_0 \\
    \end{array} \right) &= 
  \left( \begin{array}{cc}
      \cos \theta_Q & \sin \theta_Q \\
      - \sin \theta_Q & \cos \theta_Q \\
    \end{array} \right)
  \left( \begin{array}{c}
      q_L \\
      Q'
    \end{array} \right),
  \label{mixingQ}
\end{align}
where 
\begin{align}
  \cos \theta_Q &= \frac{\mu_{Q_0}}{\mu_Q},\quad \sin \theta_Q =
 \frac{\mu_{Q_Y}}{\mu_Q},
\end{align}
with $\mu_Q=\sqrt{\mu_{Q_Y}^2+\mu_{Q_0}^2}$.  Similar relations hold
for $(U_Y,U_0)$ and $(E_Y,E_0)$, with the mixing angles
$\theta_{U,E}=\cos^{-1}(\mu_{U_0,E_0}/\sqrt{\mu_{U_Y,E_Y}^2+\mu_{U_0,E_0}^2})$.
At the mass scale of the extra matters, MSSM Yukawa coupling constants
are given by
\begin{align}
  y_b (M_{\rm ex}) = &\, y_b' (M_{\rm ex}) \cos\theta_Q,
  \label{ybprime} \\
  y_\tau (M_{\rm ex}) = &\, y_\tau' (M_{\rm ex}) \cos\theta_E,
  \label{ytauprime} \\
  y_t (M_{\rm ex}) = &\, y_t' (M_{\rm ex}) \cos\theta_Q \cos\theta_U.
  \label{ytprime} 
\end{align}

In the present set up, the Yukawa structure is like that of the MSSM
as Eq.\ \eqref{W_yukawa} is obtained from the Yukawa interaction of
the MSSM by replacing $y_{t,b,\tau}\rightarrow y'_{t,b,\tau}$ and
$(q_L, t_R^c, \tau_R^c) \to (Q_Y, U_Y, E_Y)$.  Numerically, however,
such a replacement may give significant effects on the Yukawa
unification.  This is because, as shown in Eq.\ \eqref{ytprime},
$y'_t$ can be significantly larger than $y_t$ because
$\cos\theta_Q\cos\theta_U<1$.  Such an enhancement of the coupling
constant may have the following consequences:
\begin{enumerate}
\item $y_b' (M_{\rm GUT})$ is enhanced through the renormalization
  group effect while $y_\tau' (M_{\rm GUT})$ is not (see Eqs.\
  \eqref{RGEyb} and \eqref{RGEytau}).
\item $m_{H_u}^2 (M_S)$ is suppressed through the renormalization
  group effect.  This can be understood from the RGE of $m_{H_u}^2$,
  which is given by
  \begin{align}
    \frac{d}{dt} m_{H_u}^2 = 
    6  {y_t'}^2 (m_{H_u}^2 + m_{\tilde{Q}}^2 + m_{\tilde{U}}^2 + A_t^2)
    + \cdots,
    \label{mhubeta}
  \end{align}
  where only the $y'_t$-dependence of the beta-function at the
  one-loop level is shown in the above equation.  This may result in
  the enhancement of $|\mu|$ and $|\Delta_b|$.
\end{enumerate}

In order to study the effects of the Yukawa couplings of the extra
matters on the $b$-$\tau$ unification, we solve the RGEs numerically,
taking into account the effects of extra Yukawa couplings. We assume
that the threshold correction to the SUSY invariant masses of extra
matters are negligible, and that they are unified at the GUT scale; we
parameterize their boundary conditions as
\begin{align}
  &
  \mu_{Q_0} (M_{\rm GUT}) = \mu_{U_0} (M_{\rm GUT}) = \mu_{E_0} (M_{\rm GUT}) 
  \equiv \mu_{\bf 10},
  \\
  &
  \mu_{Q_Y} (M_{\rm GUT}) = \mu_{U_Y} (M_{\rm GUT}) = \mu_{E_Y} (M_{\rm GUT}) 
  \equiv X \mu_{\bf 10}.
\end{align}
Here, $X^{-1}$ is approximately equal to $\cos\theta_\Phi$ (with
$\Phi=Q,U,E$), although they slightly differ because of the
renormalization group effects from the GUT scale to the extra matter
scale.  (In our numerical analysis, we have taken into account the
effects of the renormalization group running of SUSY invariant mass
parameters.)  In addition, for $Q>M_{\rm ex}$, the scalars in the
extra matter sector have trilinear interactions.  We assume that, at the
GUT scale, the trilinear couplings are proportional to the corresponding
Yukawa coupling constants, and that the trilinear interactions above the
extra matter scale are given by
\begin{align}
  {\cal L}_{\rm trilinear}^{\rm (soft)} = 
  - A'_b H_d \tilde{Q}_Y \tilde{b}_R^c 
  - A'_\tau H_d \tilde{l}_L \tilde{E}_Y
  - A'_t H_u \tilde{Q}_Y \tilde{U}_Y,
\end{align}
with the boundary conditions $A'_b (M_{\rm GUT}) = A'_\tau (M_{\rm
  GUT}) \equiv a'_d$ and $A'_t (M_{\rm GUT}) \equiv a'_u$.
  Furthermore, the SUSY breaking mass parameters above the extra
  matter scale can be written as
\begin{align}
  {\cal L}^{\rm (soft)}_{\rm scalar\ mass} = 
  \sum_{\Phi = Q,U,E} 
  (m_{\Phi_Y}^2 | \tilde{\Phi}_Y |^2 + m_{\Phi_0}^2 | \tilde{\Phi}_0 |^2)
  + \cdots.
  \label{softmassEx}
\end{align}
Notice that, with the present choice of parameters, there is no mixing
term between $\tilde{\Phi}_Y$ and $\tilde{\Phi}_0$.  Soft SUSY
breaking parameters defined above and below the extra matter scale are
matched at $Q=M_{\rm ex}$ using Eq.\ \eqref{mixingQ} (as well as the
mixing angles for $(U_Y,U_0)$ and $(E_Y,E_0)$).

\begin{figure}
  \begin{minipage}{0.48\hsize}
    \begin{center}
      \includegraphics[width=0.98\hsize]{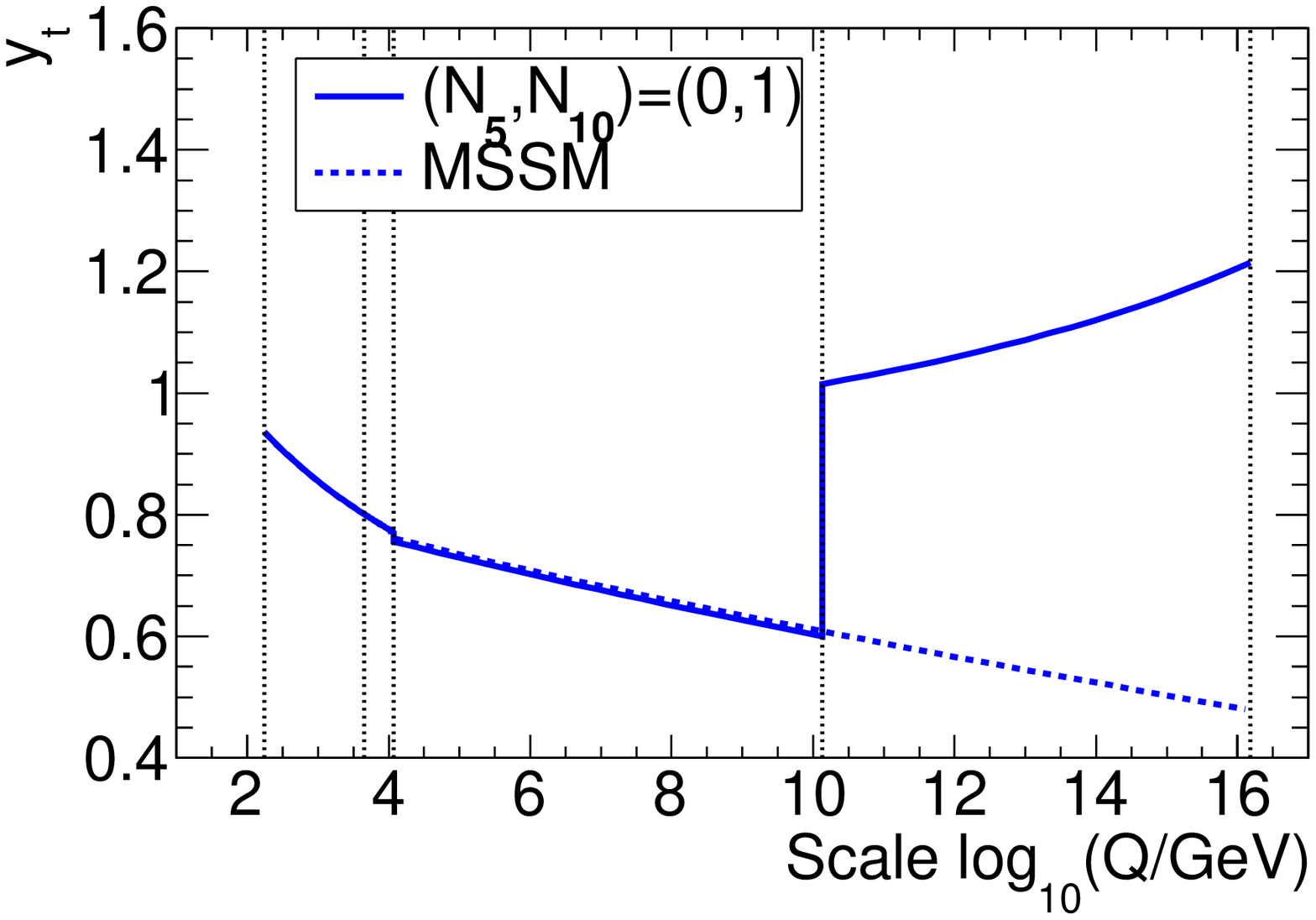}
    \end{center}
  \end{minipage}
  \begin{minipage}{0.48\hsize}
    \begin{center}
      \includegraphics[width=0.98\hsize]{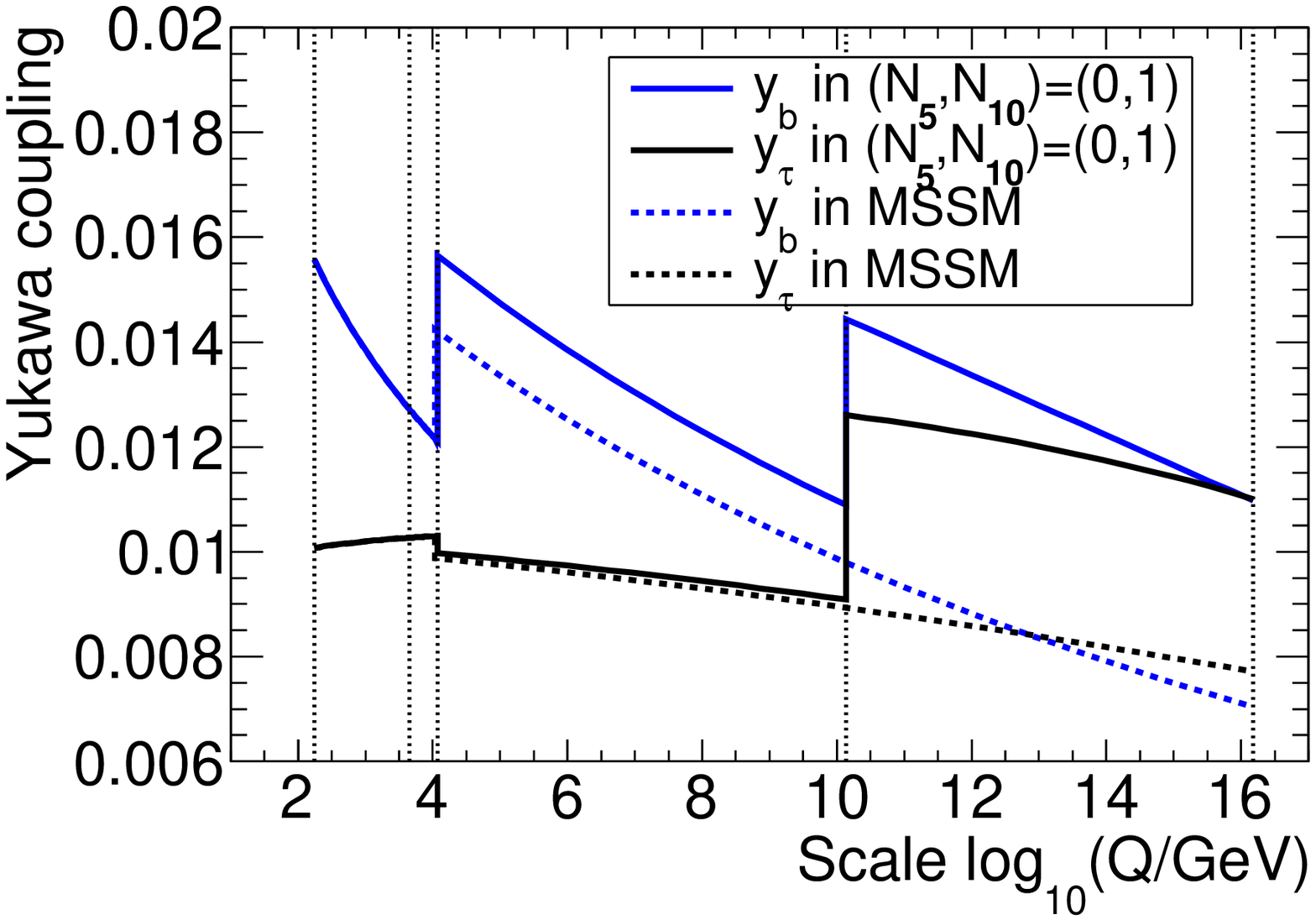}
    \end{center}
  \end{minipage}
  \caption{Runnings of $\tilde{y}_t$ and $y^{(\prime)}_t \sin \beta$ (left)
    and those of $\tilde{y}_b$, $y^{(\prime)}_b \cos \beta$,
    $\tilde{y}_\tau$ and $y^{(\prime)}_\tau \cos \beta$ (right) in the
    model with $(N_{\bf 5},N_{\bf 10}) = (0,1)$ (solid lines) and in
    the MSSM (dotted lines).  Here, we take $\tan \beta = 27$, $m_{\bf
      \bar{5}} = m_{{\bf 5}_H} = m_{{\bf \bar{5}}_H} = m_{1/2} = 3\
    {\rm TeV}$, $a_d = a_u = 0$, $\mu_{\bf 10} = 10^{10}\ {\rm GeV}$,
    and $X = 1.4$.  Here, $m_{\bf 10}$ is tuned to adjust the SM-like
    Higgs mass to be $m_h = 125.09\ {\rm GeV}$, which gives $m_{\bf
      10} = 14\ {\rm TeV}$ for the model with $(N_{\bf 5}, N_{\bf 10})
    = (0,1)$ and $m_{\bf 10} = 12\ {\rm TeV}$ for the MSSM.  The
    vertical dotted lines denote the matching scales in the model with
    extra matters: $Q = m_t$, $M_{\tilde{G}}$, $M_S$, $M_{\rm ex}$,
    and $M_{\rm GUT}$ from left to right.  For $M_S < Q < M_{\rm ex}$
    ($Q > M_{\rm ex}$), the solid lines denote $y_t \sin \beta$ ($y'_t
    \sin \beta$) or $y_f \cos \beta$ ($y'_f \cos \beta$)
    with $f=b$, or $\tau$.}
  \label{fig:YukawaRunning}
\end{figure}

We show examples of the running of Yukawa coupling constants in Fig.\
\ref{fig:YukawaRunning}, which demonstrates the possibility to make
the unification of the Yukawa coupling constants successful due to the
effects of the Yukawa interactions of extra matters.  Solid lines show
the results for the present model while dotted lines denote the result
for the MSSM as a reference.  Here, we take $\tan \beta = 27$, $m_{\bf
  \bar{5}} = m_{{\bf 5}_H} = m_{{\bf \bar{5}}_H} = m_{1/2} = 3\ {\rm
  TeV}$, $a_d = a_u = 0$, $\mu_{\bf 10} = 10^{10}\ {\rm GeV}$, and $X
= 1.4$. $m_{\bf 10}$ is used to adjust the SM-like Higgs mass in each
model to be the observed value, which gives $m_{\bf 10} = 14\ {\rm
  TeV}$ for the model with $(N_{\bf 5}, N_{\bf 10}) = (0,1)$ and $12\
{\rm TeV}$ for the MSSM.  The vertical dotted lines denote the
matching scales in the model with extra matters: from left to right,
they correspond to $Q = m_t, M_{\tilde{G}}, M_S, M_{\rm ex}$, and
$M_{\rm GUT}$, respectively.  The large ``jumps'' of solid lines in
Fig.\ \ref{fig:YukawaRunning} at $Q = M_S$ are due to the threshold
corrections, while those at $Q = M_{\rm ex}$ are mainly due to the
mixing effect represented in Eq.\ (\ref{ybprime}) and Eq.\
(\ref{ytauprime}) since we plot $y_b' \cos \beta$ and $y_\tau' \cos
\beta$ instead of $y_b \cos \beta$ and $y_\tau \cos \beta$ by solid
lines in the range $Q > M_{\rm ex}$.  We can see from the figure that
the enhancement of $|\Delta_b|$ significantly modifies the prediction
for Yukawa unification.  Together with the change in the running of
$y_b$ as we discussed before, the mixing $X > 1$ enlarges the
prediction for $R'_{b\tau}$, which is defined as
\begin{align}
  R'_{b\tau} \equiv \frac{y_b' (M_{\rm GUT})}{y_\tau' (M_{\rm GUT})}.
  \label{Rbtaunew}
\end{align}
It becomes almost $1$ in the present choice of parameters, while
$R_{b\tau} \simeq 0.91$ in the case of the MSSM with the same choice
of GUT scale boundary conditions for SUSY breaking parameters.

\begin{figure}[t]
 \begin{center}
  \includegraphics[width=90mm]{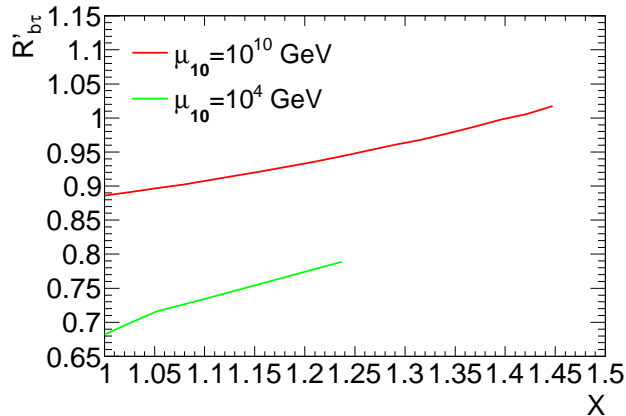}
 \end{center}
 \caption{Mixing parameter $X$ dependence of $R'_{b\tau}$ defined in
 Eq.\ (\ref{Rbtaunew}).  The parameters used are the same as Fig.\
 \ref{fig:YukawaRunning}.  $m_{\textbf{10}}$ is again used to adjust
 the SM-like Higgs mass to be $m_h = 125.09\ {\rm GeV}$ at each value
 of $X$.  The red and the green lines denote the model with
 $\mu_{\textbf{10}} = 10^{10}\ {\rm GeV}$ and $\mu_{\textbf{10}} =
 10^4\ {\rm GeV}$, respectively.}  \label{mixingVsR}
\end{figure}

In Fig.\ \ref{mixingVsR}, we show the $X$ dependence of $R'_{b\tau}$,
taking $\tan \beta = 27$, $m_{\bf \bar{5}} = m_{{\bf 5}_H} = m_{{\bf
\bar{5}}_H} = m_{1/2} = 3\ {\rm TeV}$, and $a_d = a_u = 0$.
$m_{\textbf{10}}$ is tuned for each value of $X$ to adjust the SM-like
Higgs mass to be the observed value; as a result, $m_{\textbf{10}}$
takes values in the range between $11.5\ {\rm TeV}$ and $14\ {\rm
TeV}$.  The red line shows the result for the case with
$\mu_{\textbf{10}} = 10^{10}\ {\rm GeV}$ and the green one shows that
with $\mu_{\textbf{10}} = 10^4\ {\rm GeV}$.  We can understand from
the figure that $R'_{b\tau}$ is enhanced with larger $X$.  In the
present choice of parameters with $\mu_{\textbf{10}} = 10^{10}\ {\rm
GeV}$, $R'_{b\tau}=1$ is possible.  On the other hand, for
$\mu_{\textbf{10}} = 10^4\ {\rm GeV}$, the enhancement is not so
important since in this case, the suppression of $y_b$ due to the
enhancement of the gauge coupling constants is so large (see Fig.\
\ref{MexVsR}) that it cancels the advantage of the mixing effect.
Notice that the lines in Fig.\ \ref{mixingVsR} terminate at some value
of $X$.  This is because, for larger value of $X$, $m_{\textbf{10}}$
becomes larger in order to fix the SM-like Higgs mass, which in turn
may make the right-handed sbottom mass squared or the left-handed
slepton mass squared being negative at $Q=M_S$, causing the tachyonic
sfermion problem.

\section{Summary}
\setcounter{equation}{0}

In this letter, we have studied the $b$-$\tau$ unification in SUSY
$SU(5)$ models with extra matters.  We have assumed that the extra
matters are embedded into full $SU(5)$ multiplets.  We have seen that
the extra matters may significantly affect the $b$-$\tau$ unification
in particular when the mass scale of the extra matters is much lower
than the GUT scale.

We have first considered the case where the extra matters interact
with the MSSM particles only through gauge interaction.  In such a
case, the ratio of $y_b$ and $y_\tau$ at the GUT scale, which we
called $R_{b\tau}$, becomes suppressed as the mass scale of the extra
matters becomes smaller.  This is because, with the extra matters, the
$SU(3)_C$ gauge coupling constant is enhanced at higher scale,
resulting in the suppression of the bottom Yukawa coupling constant at
the GUT scale.  The suppression of $R_{b\tau}$ has been found to be
$\sim 10\ \%$.

We have also studied the effects of the Yukawa couplings of the extra
matters with MSSM particles.  In the case we have studied, the Yukawa
couplings above the mass scale of the extra matters are effectively
enhanced, resulting in the change of the ratio of the (effective) $b$
and $\tau$ Yukawa coupling constants.  In particular, we have shown
that a simple Yukawa unification (i.e., $R'_{b\tau}=1$) can be
realized via the effects of extra matters with Yukawa interaction even
though $R_{b\tau}$ is significantly smaller than $1$ for the case
without extra matters with the same GUT scale boundary conditions.

\vspace{5mm}
\noindent {\it Acknowledgements}: This work was supported by the
Grant-in-Aid for Scientific Research C (No.26400239), and Innovative
Areas (No.16H06490).  The work of S.C. was also supported in part by
the Program for Leading Graduate Schools, MEXT, Japan.

\end{document}